# Emission spectrum of a qubit under its deep strong driving in the high-frequency dispersive regime

[1] A. P. Saiko, S. A. Markevich, *R. Fedaruk

Scientific-Practical Material Research Centre, Belarus National Academy of Sciences,

19 P. Brovka str., Minsk 220072 Belarus

*Institute of Physics, University of Szczecin, 70451, Szczecin, Poland

December 10, 2017

We study the emission spectrum of a qubit under deep strong driving in the high-frequency dispersive regime when the driving frequency and strength exceed significantly the qubit transition frequency. Closed-form expressions for the steady-state first-order field correlation function and the multiphoton emission spectrum are obtained. The spectrum comprises a series of narrow delta-like lines that stem from coherent processes and Lorenzian peaks that result from the incoherent scattering of photons. The oscillating dependence of the widths of the emission lines on the driving strength is predicted. We show how the features of this dependence are governed by the qubit dephasing and relaxation rates.

## Contents

### 1. Introduction

The semi-classical Rabi model [1] is widely used for studying and control of the resonant interaction between a classical electromagnetic field and various quantum objects (atoms, nuclear and electron spins). In this model the Hamiltonian is given by

$$H = \varepsilon s^z - 2g s^x \cos \omega t, \tag{1}$$

where $\varepsilon$ is the transition frequency of a two-level system (qubit), $g$ is the coupling strength between the qubit and the electromagnetic field with frequency $\omega$; $s^{\pm,z}$ are components of the pseudospin operator, describing the qubit states and satisfying the commutation relations: $[s^+, s^-] = 2s^z$, $[s^z, s^\pm] = \pm s^\pm$.

---

[1] e-mail: saiko@physics.by

In the past decade, theoretical and experimental studies of the matter-light interaction evolve toward the ultrastrong ($0.1 < g < \omega$) and deep strong ($g > \omega$) coupling regime where the well-known rotating wave approximation (RWA) is broken and the contribution of the antiresonant (non-RWA) terms in the coupling Hamiltonian cannot be omitted. In these regimes, the coupling strength $g$ is comparable to, or exceeds, the transition frequency $\varepsilon$ between two energy levels of the quantum system. As a result, the counter-rotating component of an electromagnetic field results in complex dynamics of the field-matter interaction (see, e. g. [2]-[7]) and makes difficulties for its analytical description.

The strong and ultrastrong regimes of light-matter interaction have been studied both theoretically and experimentally in a variety of solid state systems. The steady-state response of two-level systems, mainly superconducting qubits, under their strong continuous-wave driving have been considered [8, 9]. The strong driving of qubits has been investigated in Landau-Zener-Stuckelberg interferometry on quantum dots [10, 11] and in hybrid quantum systems composed of a superconducting qubit and a nanomechanical resonator [12] or a superconducting flux qubit and a single nitrogen-vacancy (NV) center [13]. Recently, the ultrastrong regime has been studied in time-resolved experiments with artificial atoms such as superconducting flux [14]- [16] and charge [17] qubits as well as with a single NV center in diamond [18, 19], radiation-dressed states of NV centers [20], nuclear spins [21] and mechanical driving of a single electron spin [22]. More recently, the unusual features of the deep strong regime have been reported [23]- [25].

Usually the qubit's dynamics in the strong-driving regime is described within the framework of Floquet theory in terms of quasienergies and quasienergy states [16]. The various frequency components in the observed Rabi oscillations and the Bessel-function dependence of the quasienergy difference on the driving strength have been demonstrated [15]- [17]. Dissipative and decoherence processes limiting the observation of Rabi oscillations under the ultrastrong driving regime have been considered theoretically [26, 27]. The long-time dynamics of Rabi model under the ultrastrong atom-cavity coupling for driven-dissipative scenario where the cavity is driven by a monochromatic coherent field has been studied [28]. It is interesting that the ultra- and deep strong driving field induces the transitions between the levels of the two-level system not only at its resonant or near resonant excitation, but also when the frequency of electromagnetic driving field $\omega$ is far away from resonance and exceeds significantly the qubit transition frequency $\varepsilon$ [26, 27].

In the present paper, we study the steady-state first-order field correlation function and the photon emission spectrum of the qubit under its deep strong driving in the high-frequency dispersive regime ($\omega, g \gg \varepsilon$). An analytical description is realized in the framework of the non-secular perturbation theory based on the Krylov–Bogoliubov–Mitropolsky (KBM) averaging method.

**2. Dissipative dynamics of qubit**

The master equation for the qubit interacting with a linearly polarized electromagnetic field is

$$i\frac{\partial \rho}{\partial t} = [H, \rho] + i\Lambda\rho. \qquad (2)$$

Here $H$ is denoted in (1) and $\Lambda$ is the relaxation superoperator defined as $\Lambda\rho = \frac{\gamma_{21}}{2} D[s^-]\rho + \frac{\gamma_{12}}{2} D[s^+]\rho + \frac{\eta}{2} D[s^z]\rho$, where $D[O]\rho = 2O\rho O^+ - O^+O\rho - \rho O^+O$, $\gamma_{21}$ and $\gamma_{12}$ are the rates of photon radiative processes from the excited state $|2\rangle$ of the qubit to its ground state $|1\rangle$ and vice versa, and $\eta$ is the dephasing rate. In the following, we assume that the relaxation parameters are defined phenomenologically and use the superoperator in its standard form, without its microscopic definition (see, e.g., [26], [29]-[31])).

The rapidly oscillating terms in the master equation can be eliminated in the framework of the non-secular perturbation theory by using the KBM averaging method [32]. The description of this method and its applications to studies of the dynamics of two-level systems under mono- and bichromatic driving are presented in Refs. [27, 33, 34]. In the high-frequency limit, $\varepsilon/\omega \ll 1$, and in the first order of non-secular perturbation theory the Hamiltonian $H$ is replaced by its effective counterpart:

$$H \to H_{eff} = \varepsilon J_0(a) s^z \equiv \varepsilon_q s^z, \qquad (3)$$

$a \equiv 2g/\omega$. The second order of the nonsecular perturbation theory does not yield the contribution in the effective Hamiltonian. The third order correction to the quasienergy $\varepsilon_q$ is small [27] and is neglected here. Consequently, the expression for the density matrix of a two-level system can be written as [27]: $\rho(t) = \frac{1}{2} + \left(\sigma_0 - (\sigma_0 + 1)e^{-\Gamma_{\parallel}t}\right)\left(\cos(a\sin\omega t)s^z - \frac{i}{2}\sin(a\sin\omega t)(s^+ - s^-)\right)$, where $\rho(0) = 1/2 - s^z$, $\sigma_0 = -\gamma J_0(a)/\Gamma_{\parallel}$, $\Gamma_{\parallel} = \frac{3}{4}\gamma + \frac{1}{4}\eta + \frac{1}{4}(\gamma - \eta)J_0(2a)$. (In order to applying the KBM method, we must use the values of $a \geq 1.5$.) Here we denoted $\gamma_{21}$ by $\gamma$ and assumed that at low temperatures $\gamma_{12} \approx 0$. Then we find the dipole moment of the qubit

$$D(t) = tr\left(s^+ \rho(t)\right) = \frac{i}{2}\left(\sigma_0 - (\sigma_0 + 1)e^{-\Gamma_\parallel t}\right)\sin\left(a \sin \omega t\right). \qquad (4)$$

It follows from equations for $\rho(t)$ and $D(t)$ that the qubit dynamics depends strongly on the parameter $\sigma_0$ which is equal to twice the population difference of the quasienergy states at $t \to \infty$. Due to the presence of $J_0(a)$ in the expression for $\sigma_0$, this parameter oscillates and can be negative, zero or positive. One can see that the ratio of the rates of energetic relaxation and pure dephasing influence only the amplitude of the variations in $\sigma_0$, but cannot change the sign of $\sigma_0$. Within the approximation considered here ($\omega, g \gg \varepsilon$), $\varepsilon$ does not enter the expressions for the observable values.

### 3. Correlation function and emission spectrum of qubit

Now we consider the photon emission by the qubit. The power spectrum $S(\Omega)$ of the emission is given by the Fourier transform of the first order correlation function $g^{+-}(t,\tau) = tr\left(s^+(t)s^-(t+\tau)\rho(0)\right)$:

$$S(\Omega) = \frac{1}{\pi} Re\left[\int_0^\infty d\tau e^{i\Omega\tau} \lim_{t\to\infty} g^{+-}(t,\tau)\right]. \qquad (5)$$

This function can be split into two parts

$$g^{+-}(t,\tau) = tr\left(s^+\rho(t)\right)tr\left(s^-\rho(t+\tau)\right) + tr\left(\delta s^+(t)\delta s^-(t+\tau)\rho(0)\right), \qquad (6)$$

where the dipole moment $s$ is written as a sum of an average dipole moment $\langle s \rangle$ and the instantaneous difference $\delta s$ from its average value: $s = \langle s \rangle + \delta s$. The first term in Eq. (6) is the square of the dipole average moment of Eq. (4) and describes the coherent scattering of radiation.

The second term in (6) is related to fluctuations of the dipole moment decreases with some correlation time and presents the incoherent component of the scattered radiation. Consequently, the emission spectrum from Eqs. (5) and (6) can be written as a sum of "coherent" and "incoherent" parts:

$$S(\Omega) = S_{coh}(\Omega) + S_{inc}(\Omega). \qquad (7)$$

In the steady state the dipole moment of the qubit performs undamped oscillations. Due to the high-frequency dispersive regime in the non-RWA, the driving field excites parametrically odd harmonics $(2n-1)\omega$ with $n = 1, 2, \ldots$, since $\lim_{t \to \infty} D(t) \sim \sigma_0 \sum_{n=1}^{\infty} J_{2n-1}(a) \sin((2n-1)\omega t)$ (see Eq. (4)). Consequently, at $t \to \infty$ the first term in Eq. (6) describes the coherent scattering of the radiation that occurs exactly at odd harmonics without energy dissipation. This situation is completely different from the near-resonant driving ($\omega \approx \varepsilon$), when the RWA can be used and the qubit steady state is described by only one sinusoidal oscillation at the driving frequency [35]. Using the expression $\lim_{t \to \infty} D(t)$, we find

$$S_{coh}(\Omega) = \sigma_0^2 \sum_{n=1}^{\infty} J_{2n-1}^2(a) \left[ \delta(\Omega - (2n-1)\omega) + \delta(\Omega + (2n-1)\omega) \right]. \tag{8}$$

The "coherent" part of the spectrum in Eq. (8) consists of a series of narrow delta-like peaks only at odd harmonics of the exciting field: $\omega$, $3\omega$, , $(2n-1)\omega$. The position of these peaks characterizes the frequencies of realized multiphoton quantum transitions. So, under the deep strong off-resonant driving, the qubit is parametrically excited at odd harmonics of the external field and coherently re-emits at these frequencies without energy dissipation.

The "incoherent" part of the correlation function is calculated by invoking the quantum regression theorem and can be written as

$$S_{inc}(\Omega) = \frac{1}{4\pi} \left(1 + \sigma_0 J_0(a)\right) \Gamma_\perp \left\{ \frac{1 + J_0(a)}{\left(\Omega - \varepsilon_q\right)^2 + \Gamma_\perp^2} + \right.$$
$$\left. + \sum_{n=1}^{\infty} J_{2n}(a) \left( \frac{1}{\left(\Omega - 2n\omega - \varepsilon_q\right)^2 + \Gamma_\perp^2} + \frac{1}{\left(\Omega + 2n\omega - \varepsilon_q\right)^2 + \Gamma_\perp^2} \right) \right\}, \tag{9}$$

where the following "transverse" relaxation parameter appears:

$$\Gamma_\perp = \frac{5}{8}\gamma + \frac{3}{8}\eta + \frac{1}{8}(\eta - \gamma)J_0(2a). \tag{10}$$

According to Eq. (9), the "incoherent" part of the multiphoton emission spectrum consists of a series of Lorentzian peaks. A low-frequency peak is centered at the frequency equal to the quasienergy

$\Omega^{(0)} = \varepsilon_q$. Other peaks are observed at frequencies $\Omega^{(n)} = 2n\omega + \varepsilon_q$, where $n = 1, 2, \ldots$. We see that the incoherent photons are emitted at the frequency of the qubit quasienergy $\varepsilon_q$ as well as at even harmonics of the exciting field. These harmonics are shifted towards higher (lower) frequencies by the value of $|\varepsilon_q|$ ($-|\varepsilon_q|$). Relative intensities of the low-frequency and other peaks are given by $1 + J_0(a)$ and $J_{2n}(a)$, respectively. All these spectral lines are broadened by coupling with the environment and have the same width $2\Gamma_\perp$. The central frequencies, widths and relative intensities of the peaks depend on the driving strength. It follows from Eq. (10) that the relaxation rate $\Gamma_\perp$ oscillates in accordance with the Bessel-function dependence $J_0(2a)$. As a result, the widths of the Lorentzian lines in the "incoherent" emission spectrum oscillate with increasing the driving strength. Like the case of the relaxation rate $\Gamma_\parallel$ (see [27]), the features of these oscillations are determined by the ratio of the rates of energetic relaxation $\gamma$ and pure dephasing $\eta$. If these rates are equal to each other ($\gamma = \eta$), the relaxation rate of the population difference does not dependent on the driving strength. At $\gamma > \eta$, the variations of $\Gamma_\perp$ are inverted in comparison with the case $\gamma < \eta$, because the coefficient before the Bessel function in Eq. (10) changes its sign.

Thus, in the high-frequency dispersive regime ($\omega, g \gg \varepsilon$), the rotating $s^+ e^{-i\omega t}$ and counter-rotating $s^+ e^{i\omega t}$ terms in the Hamiltonian (1) play an equivalent role. It leads to the parametric excitation of multiphoton transitions and the new features of the emission spectrum. Therefore, the coherent and incoherent emission spectra differ radically from those obtained in the RWA, when the emission spectrum consists of the Mollow triplet and the coherent $\delta$-like peak at the driving frequency [35].

For numerical calculations we use the following system of equations for the correlation functions $g^{+-}(t, \tau)$, $g^{+z}(t, \tau) = tr\left(s^+(t) s^z(t+\tau) \rho(0)\right)$, $g^{++}(t, \tau) = tr\left(s^+(t) s^+(t+\tau) \rho(0)\right)$:

$$\frac{\partial}{\partial \tau} g^{+-}(t, \tau) = -i\varepsilon g^{+-}(t, \tau) - ia\omega \cos(\omega \tau) g^{+z}(t, \tau) - \gamma_\perp g^{+-}(t, \tau), \qquad (11)$$

$$\frac{\partial}{\partial \tau} g^{+z}(t, \tau) = (i/2) a\omega \cos(\omega \tau) \left(g^{++}(t, \tau) - g^{+-}(t, \tau)\right) - \gamma_\parallel \left(g^{+z}(t, \tau) + 1/2\right),$$

$$\frac{\partial}{\partial \tau} g^{++}(t,\tau) = i\varepsilon g^{++}(t,\tau) + ia\omega \cos(\omega\tau) g^{+z}(t,\tau) - \gamma_\perp g^{++}(t,\tau)$$

with initial conditions

$$\begin{aligned} g^{+-}(t,\tau=0) &= 1/2 + tr\left(s^z \rho(t)\right), \\ g^{++}(t,\tau=0) &= 0, \\ g^{+z}(t,\tau=0) &= -tr\left(s^+ \rho(t)\right). \end{aligned} \quad (12)$$

The rates of the transverse and longitudinal relaxations in Eq. (11) are defined as $\gamma_\perp = (\gamma+\eta)/2$ and $\gamma_\| = \gamma$. In Fig. 1 we plot the emission spectra $S(\Omega)$ which were calculated numerically using Eqs. (5), (11) and (12) and analytically using the obtained formulas (6) - (9) for different values of the driving strength $a$. These spectra correspond to three different regimes of the qubit's dynamics [27], when the values of $\sigma_0$ are negative (Fig. 1a and b), positive (Fig. 1d and e) and zero (Fig. 1c). The narrow delta-like lines at odd harmonics $(2n-1)\omega$, where $n=1,2,...$, show the "coherent" part of the emission spectrum, $S_{coh}(\Omega)$. The Lorentzian lines at the frequencies $\varepsilon_q$ and $\Omega^{(n)} = 2n\omega + \varepsilon_q$ present the "incoherent" part, $S_{inc}(\Omega)$. There is reasonable agreement between the numerical and analytical results. We illustrate that the driving field can invert some Lorentzian lines as their intensities are proportional to the functions $J_{2n}(a)$, which can change their sign with changing the driving strength $a$. It means that at the certain frequencies the absorption of photons by the qubit can take place instead their emission.

From the numerically calculated emission spectrum we can find the quasienergy $\varepsilon_q$ and compare with its analytical expression $\varepsilon_q = \varepsilon J_0(a)$ in Eq. (3). Fig. 2 shows the quasienergy $\varepsilon_q$ as a function of the driving strength $a$. The analytical results are in a good agreement with the numerical calculations of the frequency shift, $\Omega^{(n)} - 2n\omega = \varepsilon_q$, of the "incoherent" peaks in the emission spectrum relative to even harmonics.

To our knowledge, the obtained emission spectrum $S(\Omega)$, comprising a series of coherent delta-like peaks at odd harmonics, $(2n-1)\omega$, of the driving field and incoherent relaxation-broadened peaks at even harmonics, $2n\omega + \varepsilon_q$, shifted by the quasienergy value, has not been described in the literature. As an unexpected result, we find the inversion of some incoherent lines at the certain values of the driving strength. We hope that these new features of the emission spectrum, originating from the strong effect of the counter-rotating terms, will stimulate future experiments.

## 4. Conclusions

We have found that under deep strong driving of the qubit in the high-frequency dispersive regime the multiphoton emission spectrum consists of the coherent (the narrow delta-like lines at odd harmonics of the driving field) and incoherent (the Lorentzian lines at frequencies equal to the quasienergy and the sum of the quasienergy and even harmonics of the driving field) parts. It is shown that the widths of emission lines have the Bessel-function-like dependences on the driving strength. Moreover, the driving field can invert some Lorentzian lines of the emission spectrum and at the certain frequencies the absorption of photons by the qubit can take place instead their emission. The obtained new features of emission spectrum are fundamental and important to physics of open quantum systems at deep strong far-off-resonant driving as well as for potential practical applications, including nonlinear spectroscopy and multi-frequency processing.

The work was supported by Belarusian Republican Foundation for Fundamental Research (Grant F16MS-013) and by State Programm of Scientific Investigations "Physical material science, new materials and technologies", 2016-2020.

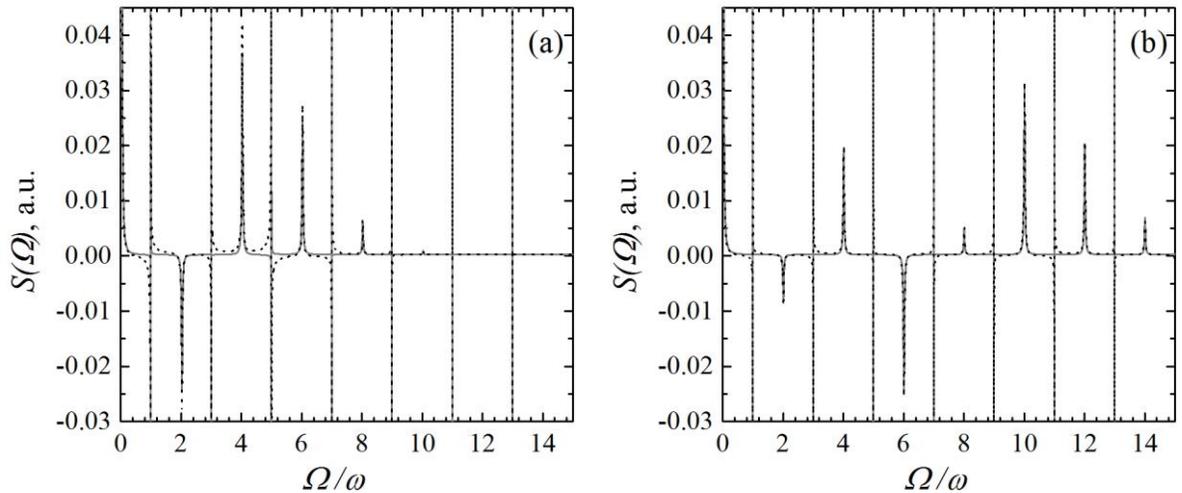

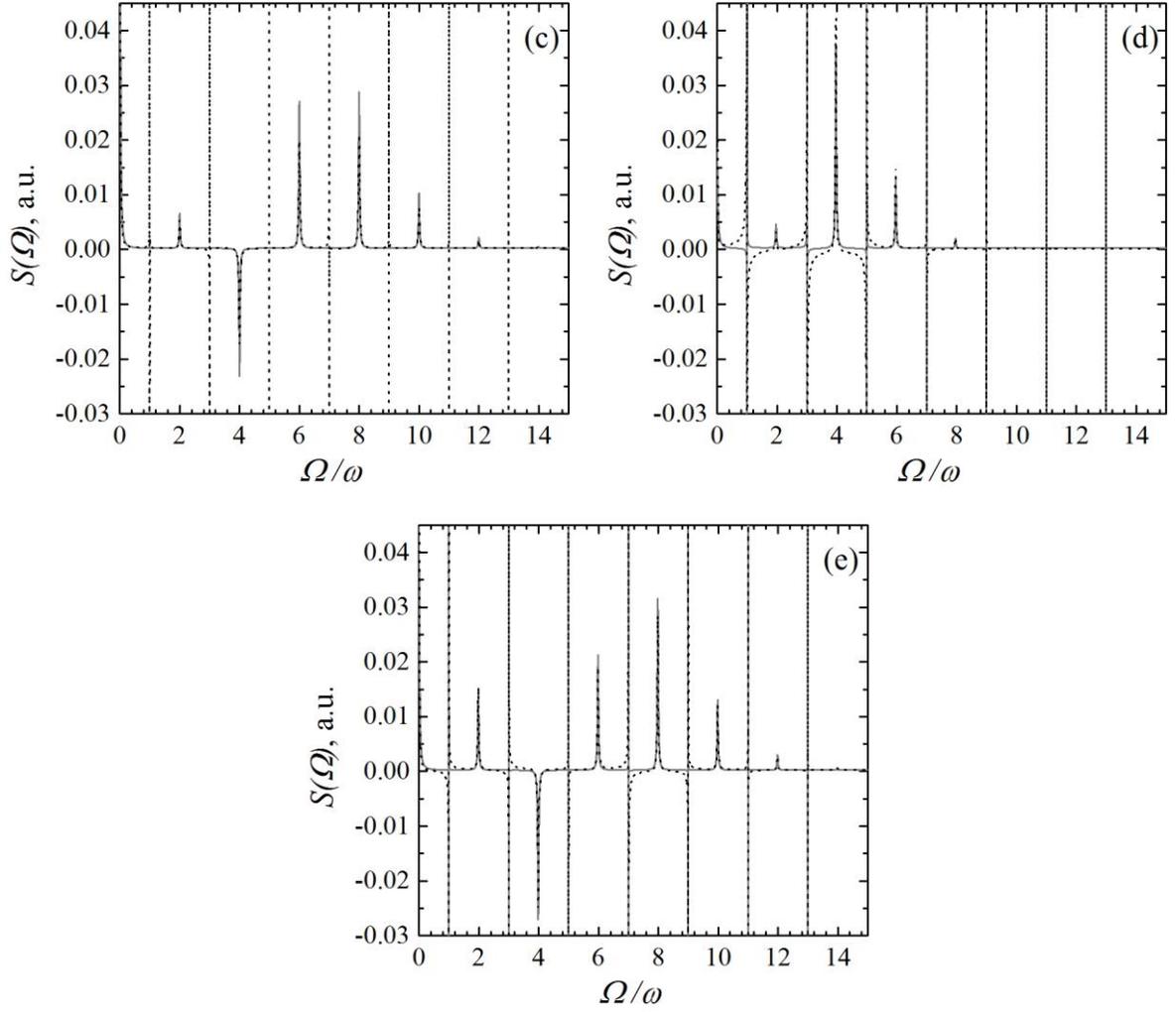

Fig 1. The emission spectra for different values of the driving strength at $\gamma/\omega = 0.03$ and $\eta = 0$.
(a) $a = 6.0$, $\sigma_0 = -0.198$ (b) $a = 12.0$, $\sigma_0 = -0.065$ (c) $a = 8.6$, $\sigma_0 = 0$ (d) $a = 5.0, \sigma_0 = 0.258$
(e) $a = 9.0$, $\sigma_0 = 0.12$. The dotted lines show the numerical results.

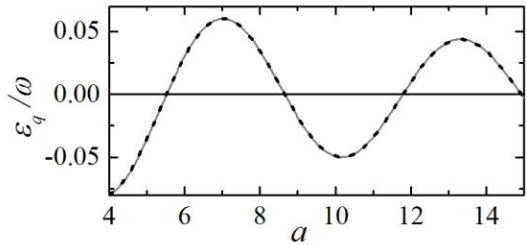

Fig. 2. The quasienergy $\varepsilon_q$ as a function of the driving strength at $\gamma/\omega = 0.03$ and $\eta = 0$. The dashed line presents the numerical results.